\documentclass[11pt]{article}
\usepackage{amssymb}
\usepackage{pst-all}
\usepackage{pstricks}
\usepackage{pstcol,pst-fill,pst-grad}
\usepackage{amsfonts}
\usepackage{graphicx}
\usepackage{amsmath}
\usepackage[normalem]{ulem}
\usepackage{multicol}
\usepackage{tikz}
\usepackage{enumerate}

\usetikzlibrary{matrix}
\RequirePackage{mathrsfs} \RequirePackage[sc]{mathpazo}
\RequirePackage{wasysym} \RequirePackage{setspace}
\makeatletter \@addtoreset{equation}{section}

\newcommand{\be}{\begin{equation}}
\newcommand{\ee}{\end{equation}}
\newcommand{\bea}{\begin{eqnarray}}
\newcommand{\eea}{\end{eqnarray}}
\input{tcilatex}
\begin{document}

\date{}
\title{\textbf{\ Probing New Physics From TXS 0506+056 Blazar Neutrinos}}
\author{ Adil Belhaj$^1$ \thanks{%
belhaj@unizar.es}, Salah Eddine Ennadifi$^2$ \thanks{%
ennadifis@gmail.com} \\
\\
{\small $^1$ ERPTM,  Physics department, Polydisciplinary Faculty,   Sultan Moulay Slimane
University }\\
{\small  B\'eni Mellal, Morocco } \\
{\small $^{2}$ Independent Researcher, Rabat, Morocco}}
\maketitle

\begin{abstract}
Supported by the recent detection of a TXS 0506+056 blazar neutrino
with an energy in the TeV-PeV range detected by the IceCube
experiment, a probe of a new physics scale $\Lambda_{\mbox{NP}}$
related to high-energy cosmic neutrinos is provided. At Standard
Model  energies, the effect of the underlying physics ${\Lambda
}_{\mbox{NP}}$ is first  investigated. Then, the recorded IceCube-170922A
data is used to approach the scale of such a new physics which lies   under the Grand Unified Theory scale $ {\Lambda
}_{\mbox{NP}}\leq 10^{25}eV$.


\textbf{PACs}: 11.10.-z; 12.60.-i; 98.54.Cm.
\end{abstract}

\newpage

\section{Introduction}

Recently, with the progressive interest in neutrino physics, the
quest of neutrino properties still receives massive impetus. There  have been relevant efforts related to
such a physics.
Precisely, many investigations have been conducted to understand the
nature and the origin of the existing mass, as well as the
propagation behavior of such  puzzling particles \cite{1,2}. Long
baseline experiments are not the only way to study new physics (NP)
related to neutrinos beyond the standard model (SM) of particle
physics \cite{3,4,5}. It has been remarked that fluxes of neutrinos
are predicted to be potentially produced by a wide variety of cosmic
objects \cite{6,7}. Among these prospective sources, however, only
neutrinos from the Sun and Supernova 1987A have been observed and
used in order to deal with neutrino properties and capabilities. It
has   been suggested that high-energy neutrinos (HEN)
from distant astrophysical objects, traveling along the
intergalactic way towards Earth, open a new window to identify their
sources and suspect the NP effect from the collected data.

More recently, a HEN in the $TeV$-$PeV$ range has been detected by
the IceCube telescope and has been found to be associated with a
blazar, which is a quasar with a relativistic jet, i.e. an Active
Galactic Nuclei (AGN) \cite{8,9,10,11}. Concretely, the revealed
IceCube-170922A event has been announced with a considerable
probability $\geqslant 50\%$ of being a truly astrophysical
neutrino. In this context, the most probable energy has been
found to be around $\sim 300TeV$ with a high confidence level $C.L$ $%
\geqslant 90\%$. The best-precised retraced direction highly matched
with
the BL Lac object TXS 0506+0561, being at $0.1^{%
{{}^\circ}%
}$ from its celestial position. This blazar, with a redshift of
$z=0.3365\pm 0.0010$, is about $4$ billion light years from Earth.
The activity of TXS 0506+056 in all electromagnetic bands has been
shown by the extensive follow-up observations by other detectors,
i.e. Fermi-LAT, MAGIC and ANTARES \cite{12,13}. The observation of
such ultra-high energy cosmic rays and astrophysical neutrinos comes
up with a new road that could unveil some long standing particle
physics and cosmological problems. Indeed, the corresponding
measurements offer a unique opportunity to explore a NP going beyond
the already probed scales, including the one of the SM.

The objective of this paper is to contribute to these triumphing
achievements by investigating the NP related to HENs $\sim
TeV$-$PeV$. In particular, we exploit the TXS 0506+056 neutrino
along with the known SM data to probe the scale $\Lambda _{NP}$ of
the underlying NP.

The paper is organized as follows. In the section 2, we give a
concise review on the physics of cosmic HEN including sources and
detections. In section 3, we present the motivations for NP related
to neutrinos. In section 4, we approach the scale $\Lambda _{NP}$
from the TXS 0506+056 neutrino physical constraints. The last
section is devoted to discussions and perspectives.

\section{High energy cosmic neutrinos}

In this section, we give a concise review on HEN. More details on
such elusive particles can be found in relevant references. It is
known that neutrinos correspond to hadronic processes in the
Universe. Concretely, such objects could point to the origin of
cosmic rays. Generally, high-energy particles $E_{particle}\gg 1TeV$
can be produced by either particle acceleration in terrestrial
laboratories or by acceleration of protons in certain astrophysical
sources. It is has been noticed that the sources of HEN can be then
distinguished either by the type of neutrino production or by the
way of detection. Possible cosmic accelerators are associated with
observations of high-energy photons. In astrophysical sources such
as young supernova residues, binary black hole systems, and the
interactions of the produced protons with targets $X$ in the
interstellar medium yield pions and other particles according to the
following processes
\begin{equation}
p+X\rightarrow \pi ^{\pm ,0}+...  \label{eq1}
\end{equation}%
where the produced neutral pions decay into photons
\begin{equation}
\pi ^{0}\rightarrow \gamma +...,\text{ \ }  \label{eq2}
\end{equation}%
and the charged pions decay into neutrinos $\nu _{\ell }$ and
related leptons $\ell $ as follows
\begin{equation}
\pi ^{\pm }\rightarrow \ell +\nu _{\ell }.\text{\ }  \label{eq3}
\end{equation}%
It turns out that  the  relevant suited sources of HEN are
gamma ray bursts. The intense radiation released in terms of
energetic photons interacts with nucleons on their way through the
surrounding sphere generating neutrinos through the processes
\begin{equation}
\gamma +p\rightarrow n+\pi ^{+}...\rightarrow \mu ^{+}+\nu _{\mu
}+...\rightarrow e^{+}+\nu _{e}+\overline{\nu }_{\mu }+...
\label{eq4}
\end{equation}%
The production of cosmic HEN by gamma ray bursts is then indicated
by the observation of gamma ray burst-neutrino coincidence. In
addition to gamma ray bursts, active galactic nuclei are the
brightest sources in the Universe. The two jets of accelerated
particles near the center emerge in opposite directions
perpendicular to the disc of the active galactic nuclei are dumped
on the surrounding matter. Although we do not know exactly where or
how these particles are accelerated to these extreme energies,
reasonably we could  say that the gravitational energy released in the
accretion of the surrounding matter can engender  the acceleration of
such particles. In this way, HENs are produced by the interactions
of these particles at various stages from the source to cosmic voids
(\ref{eq4}).

HENs interact predominantly with matter in deep inelastic scattering
of  nucleons $N$ by means of neutral and charged currents of the
weak interaction according to the SU(2) gauge theory. It is recalled
that
 SU(2)  is the symmetry group of such an interaction. In
particular, they
scatters of
 quarks in the target nucleus by the exchange of a $Z^{0}$ or $%
W^{\pm }$ weak gauge field bosons. In the $Z^{0}$ interaction, the
neutrino state $\nu _{\ell }$ is left intact as
\begin{equation}
\nu _{\ell }+N\overset{Z^{0}}{\rightarrow }\nu _{\ell }+...\text{\
,} \label{eq5}
\end{equation}%
while in the $W^{\pm }$ interaction, however, a charged lepton $\ell
$ is produced sharing the initial neutrino flavor $\nu _{\ell }$ as
\begin{equation}
\nu _{\ell }+N\overset{W^{\pm }}{\rightarrow }\ell +....
\label{eq6}
\end{equation}%
These both scatterings could be schematized in the figure 1.

\begin{center}
\begin{figure}[th]
\begin{center}
{\includegraphics[width=10cm]{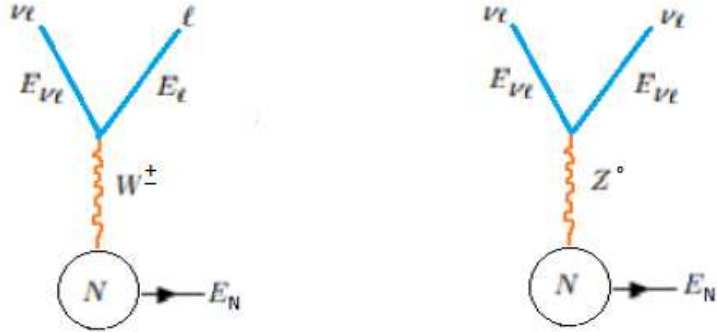}} \vspace*{-.2cm}
\end{center}
\caption{Diagrams for charged (left) and neutral (right) current
neutrino-nucleon interaction. Conservation of energy requires
$E_{\nu }=E_{\ell }+E_{N}$. Time flow is from left to right.}
\label{fig1:fig2x}
\end{figure}
\end{center}

To observe cosmic HENs, Immense particle detectors are needed.
Indeed, since a few decades ago, it has been confirmed that a
detector of kilometer-scale was required \cite{14,15}. Among the
variety of methods used to detect HENs, an effective detector is the
one exploiting the Cherenkov
radiation$\footnote{%
A transparent medium instrumented with photomultipliers that
transform the Cherenkov light into electrical signals using the
photoelectric effect.}$, based on the observation of Cherenkov light
released by the superluminal (phase velocity) propagation of
secondary charged particles produced by HEN interactions with the
detector's medium (Fig.1). Analysis of the Cherenkov light leads to
a reconstruction of theirs patterns and gives information on their
arrival directions including primordial characteristics. For
instance, an easily identified class of Cherenkov emission is based
on the passage of long-lived muons $\mu $ through the detector that
can be produced in the charged-current interactions of muon
neutrinos $\nu _{\mu }$. It follows from (\ref{eq6}) that one can
consider elongated \textit{tracks }unlike the
energetic electrons $e$ or taus $\tau $\ of the electron and tau neutrino $%
\nu _{e}$ and  $\nu _{\tau }$ interactions due to electrons rapid
scattering and taus short lifetime. For the sake of the large
background of atmospheric muons, the observation of muon neutrinos
is limited to upcoming muon tracks by neutrinos that have passed
through the Earth, used as a filter. In particular, this can help to
identify neutrinos that interact outside the detector, which limits
the neutrino view to a single flavor and to $1/2$ of the sky. In the
actual immense neutrino observatories, especially IceCube and
ANTARES, the instrumented transparent medium is  based  either on ice or
water \cite{8,13}.

\section{New Physics related to neutrinos}

According to  the SM physics, neutrinos are massless and
left-handed. It is recalled that SM is an $\mbox{SU(3)}_C\otimes
\mbox{SU(2)}_L\otimes \mbox{U(1)}_Y$ gauge theory, which describes
three generations of quarks and leptons
governed  by the strong, weak and electromagnetic fundamental interactions \cite%
{3,4,5}. This theory has been considered as a particular model of a
quantum field theory dealing with matter and forces as quantum
fields whose excitations are associated with fundamental particles
including neutrinos. Due to its limited predictions, the most
exciting physical task is, however, to search for a NP from HENs.
The leading candidate for such a NP could be related to the neutrino
physics. For the generation of the mass of neutrinos, new
interactions going beyond the SM have to be involved in such an
investigation direction. In the SM with left-handed, two-component
Weyl field neutrinos $\nu _{\ell L}$, the term of the neutrino mass
can be generated only by a beyond-SM mechanism. A general powerful
method which allows to describe effects  corresponding to such a
mechanism in  the electroweak region is the effective Lagrangian way
\cite{16,17}. This effective model  is  based on a Lagrangian formed
by the SM fields being an $ \mbox{SU(2)}_L\otimes \mbox{U(1)}_Y$
invariant and nonrenormalizable. It could be expressed in the
following form

\begin{equation}
\zeta _{4+n}^{eff}=\dsum\limits_{n\geqslant
0}y_{f}\frac{O_{4+n}}{\Lambda ^{n}}+h.c.,  \label{eq7}
\end{equation}%
describing either exotic couplings or corrections to the low-energy
SM interactions. The number $n$ appearing  in (\ref{eq7}) is a
positive integer that will be involved from now in the present study. The operator
$O_{4+n}$ is a $
\mbox{SU(2)}_L\otimes \mbox{U(1)}_Y$ invariant term with mass dimension $M^{4+n}$, and the scale $%
\Lambda $ characterizes a NP beyond the SM. In spite of the fact
that the available amount of experimental information is still
insufficient to incite for a particular NP candidate, there are,
certainly, various paths to select a particular NP from others.
Because the obvious way to implement neutrino masses to the SM leads
to the loss of renormalizability, only neutrino mass terms
consistent with the SM symmetries are allowed like
\begin{equation}
\zeta _{\nu _{mass}}^{eff}=\dsum\limits_{n\geqslant 1}y_{f}\frac{O_{4+n}}{%
\Lambda ^{n}}+h.c..  \label{eq8}
\end{equation}%
Though the possibility of Majorana mass terms is allowed because of
the electrical neutrality, such terms are prohibited in the SM and
the electroweak gauge symmetry $\mbox{SU(2)}_{L}\otimes
\mbox{U(1)}_{Y}$ forbids the ordinary four-dimensional ($n=0$)
Yukawa interactions. The task that we are after pushes one to seek
for irrelevant terms. For instance, this can be done by using
invariant higher mass dimensional effective terms ($n\geqslant 1 $)
made out of the SM fields, parameterizing the effect of new-physics
degrees of freedom on the low-energy theory. Forgeting about spinor,
gauge and flavor field indices, the lowest order term $O_{4+1}$ is
$(\ell h)^{2}$ with $h$ stand for the SM Higgs field \cite{18}. It
is recalled that the
fields $\ell ^{T}=\left( \nu _{\ell }\text{, }\ell \right) $ and $%
h^{T}=\left( h^{+}\text{, }h^{0}\right) $ are $\mbox{SU(2)}_{L}$
doublets with a nonzero $\mbox{U(1)}_{Y}$ hypercharge, and are
singlet in the color space associated with the $\mbox{SU(3)}_{C}$
gauge theory. The matter that we would like to address here is what
are the higher dimensional ($n\geqslant 1$) effective mass terms
when the lowest dimensional term $O_{4+1}$ term is not available.
This gives the possibility to envisage higher dimensional terms.
Actually, by employing the invariant Higgs combination $(h^{\dag
}h)$, the effective
neutrino mass term can be generally written at any mass dimension ($%
n\geqslant 5$) in the unique form
\begin{equation}
O_{2n+5}=(\ell h)^{2}(h^{\dag }h)^{n},\text{ \ \ }2n+5=5,\text{ }7,\text{ }%
9,...  \label{eq9}
\end{equation}%
where now the subscript $2n+5$ is the corresponding dimensions of
the neutrino mass term. A close inspection shows that underlying
theories could then be discerned by their different inputs in the
possible neutrino term dimensions (\ref{eq9}). The manifestation of
such a term is more general and would be expected to appear in any
high-scale theory in which the lepton number is violated. By inverse
powers of the scale $\Lambda _{NP}$ of the underlying NP, this term
could be dimensionally reduced. Thus, the new coupling correction
(\ref{eq8}) to the low-energy SM interactions reads as
\begin{equation}
\zeta _{\nu _{mass}}^{eff}=y_{\nu }\frac{(\ell h)^{2}(h^{\dag }h)^{n}}{%
\Lambda _{NP}^{2n+1}}.\text{ \ }  \label{eq10}
\end{equation}%
Here, the factor $y_{\nu }/\Lambda _{NP}^{2n+1}$ represents the
effective Yukawa coupling constant. After breaking of the
electroweak symmetry by the
Higg vev $\left\langle h\right\rangle $ down to the electromagnetism $%
\mbox{SU(2})_{L}\otimes \mbox{U(1)}_{Y}\overset{\left\langle h\right\rangle }%
{\rightarrow }\mbox{U(1)}_{Q_{em}}$, the above term gives rise to
the follwing suppressed Majorana neutrino mass
\begin{equation}
m_{\nu _{mass}}\simeq y_{\nu }\frac{\left\langle h\right\rangle ^{2n+2}}{%
\Lambda _{NP}^{2n+1}}\text{\ .}  \label{eq11}
\end{equation}%
It is worth mentioning that if the scale $\Lambda _{NP}$ is very
high, such neutrino mass could be thought of as induced by some
gravitational Planck scale effects \cite{19}. In this way, with
$y_{\nu }\sim 1$, one has
\begin{equation}
\Lambda _{NP}\sim \Lambda _{Planck}.  \label{eq12}
\end{equation}%
If it is the case, however, there is no eventuality of a direct
observation
of the relative NP. This means that the corresponding suppressed neutrino masses $%
m_{\nu }\ll eV$ might be too small to explain the known neutrino
masses. Subsequently, not very new scales of physics $\Lambda
_{NP}<\Lambda _{Planck} $ occur to induce the wished mass to
neutrinos. Despite of the absence of a fully consistent theory at
hand, an open window to the NP scale $\Lambda _{NP}$ could be
possible by considering its effect that might be felt at low
energies from the neutrino kinematic. The presence of the higher
mass scale $\Lambda _{NP}$ behind the neutrino mass is quiet
foreseeable to manifest at low energy in the neutrino sector. In
this approach, the neutrino behavior, unlike the other particles,
undergoes the effect of the relative $NP$. This could be examined at
the neutrino propagation where its energy $E_{\nu }$ depends, in
addition to the momentum $p_{\nu }$, on the mass scale $\Lambda
_{NP}$. With the the presence of  mass appearing in  (\ref{eq11}), the
neutrino dispersion relation is given by  \footnote{%
Natural units $c=\hbar =1$ are used.}.

\begin{equation}
E_{\nu }\simeq \sqrt{p_{\nu }^{2}+y_{\nu }^{2}\frac{\left\langle
h\right\rangle ^{4n+4}}{\Lambda _{NP}^{4n+2}}}\text{\ .}
\label{eq13}
\end{equation}%
Using the fact of the littleness of the neutrino mass compared to
its energy $y_{\nu }\left\langle h\right\rangle ^{2n+2}/\Lambda
_{NP}^{2n+1}\ll E_{\nu } $, the corresponding neutrino velocity is

\begin{equation}
v_{\nu }\left( E_{\nu }\right) \simeq 1-\frac{y_{\nu
}^{2}\left\langle h\right\rangle ^{4n+4}}{2E_{\nu }^{2}\Lambda
_{NP}^{4n+2}}\text{\ ,} \label{eq14}
\end{equation}%
where the energy-dependent term $\sim y_{\nu }^{2}\left\langle
h\right\rangle ^{4n+4}/2E_{\nu }^{2}\Lambda _{NP}^{4n+2}$ is now the
retardation, from the speed of light, effect undergone by the
propagating neutrino. This retardation effect could be experienced
by the neutrino time
delay $\delta t_{\nu -c}$ with respect to a light ray propagating at speed $%
c=1$ and emitted by the same source at a distance $d$ from the
detector like
\begin{equation}
\delta t_{\nu -c}\left( E_{\nu }\right) \simeq d\frac{y_{\nu
}^{2}\left\langle h\right\rangle ^{4n+4}}{2E_{\nu }^{2}\Lambda _{NP}^{4n+2}}%
\text{\ .}  \label{eq15}
\end{equation}%
In case of terrestrial long baseline experiments, i.e. of the order
of the
Earth radius $d^{Terrestrial}\sim 10^{6}m$, and if the underlying scale $%
M_{NP}$ is huge, the corresponding time delay values might be
extremely small $\delta t_{\nu -c}\ll 1s$ making it hardly
detectable. Thus, an adequate experimental time resolution is then
required to detect such suppressed effects. In order to probe these
conjectured NP effects which have to be discerned from those of
conventional media, distant pulsed sources, i.e. from other galaxies
$d^{Galactic}\ggg d^{Terrestrial}$ emitting HENs constitute an
insurance premium. It is expected that neutrinos with their
low-interaction cross section may provide the best scope for cosmic
sources of HEN from the largest distances.

\section{TXS 0506+056 physical constraints}

It is known that prominent candidate sources of HENs emission are
blazars. They are a class of AGN with strong relativistic jets
directed near to our line of sight with highly variable (on
time-scales from 1sec to 1year) electromagnetic emission observed.
The recent detection of a HEN of $E_{\nu }^{TXS}\sim 300TeV$,
IceCube-170922A, in direction and time coincidence with a
high-energy gamma-ray $E_{\gamma }^{TXS}\sim GeV$ flare from the
blazar TXS 0506+056 \cite{8,9,10,11}, qualifies the latter as the
first identifiable astrophysical source of HEN flux. In principle,
the huge distance of the HEN and gamma rays source TXS 0506+056,
which is at about

\begin{equation}
d^{TXS}\simeq 4.10^{9}l.y\sim 10^{24}m\text{\ ,}  \label{eq16}
\end{equation}%
from Earth, crossing neutrino including gamma rays data from this
distant source may offer an opportunity to observe neutrinos over a
baseline that is roughly $\sim 10^{20}$ times longer than that
traveled by solar neutrinos. The event recorded with such a far HEN
emitting source provides a powerful tool to suspect scenarios of the
NP scale $\sim M_{NP}$. \ Using the long travel time of the HEN as
\begin{equation}
t_{\nu }^{TXS}\simeq d^{TXS}+\delta t_{\nu -c}^{TXS}\left( E_{\nu
}^{TXS}\right) =\left( 1+\frac{y_{\nu }^{2}\left\langle
h\right\rangle ^{4n+4}}{2\left( E_{\nu }^{TXS}\right) ^{2}\Lambda
_{NP}^{4n+2}}\right) \sim 4.10^{9}y\text{,}  \label{eq17}
\end{equation}%
corresponding to the long intergalactic path (\ref{eq16}) traveled
by the HEN, the resulting time delay (\ref{eq15}) might be slightly
amplified. In particular, it is given by
\begin{equation}
\delta t_{\nu -c}^{TXS}\left( E_{\nu }^{TXS}\right)
=d^{TXS}\frac{y_{\nu }^{2}\left\langle h\right\rangle
^{4n+4}}{2\left( E_{\nu }^{TXS}\right) ^{2}\Lambda
_{NP}^{4n+2}}\gtrsim 10^{-56n-56}s\text{,}  \label{eq18}
\end{equation}%
where we have used the bound $\Lambda _{NP}\lesssim \Lambda
_{Planck}\sim 10^{28}eV$, the orders $y_{\nu }^{2}\sim 1$,
$\left\langle h\right\rangle \sim 10^{2}GeV$, the HEN energy $E_{\nu
}^{TXS}\sim 300TeV$, and the light travel time $d^{TXS}$ ($c=1$)
from TXS 0506+056. With these data, the relation (\ref{eq18}) helps
one to express and bound the underlying NP scale. The finding result is given  by
\begin{equation}
\Lambda _{NP}\simeq \sqrt[4n+2]{d^{TXS}\frac{y_{\nu
}^{2}\left\langle h\right\rangle ^{4n+4}}{2\left( E_{\nu
}^{TXS}\right) ^{2}\delta t_{\nu -c}^{TXS}}}\lesssim
10^{\frac{50n+28}{2n+1} }eV\text{.}  \label{eq19}
\end{equation}%
This relation naturally allows one to probe the possible range of
the NP scale by expanding the exponent $n$ given in (\ref{eq9}) to
its extreme values as illustrated in figure 2.
\begin{center}
\begin{figure}[th]
\begin{center}
{\includegraphics[width=10cm]{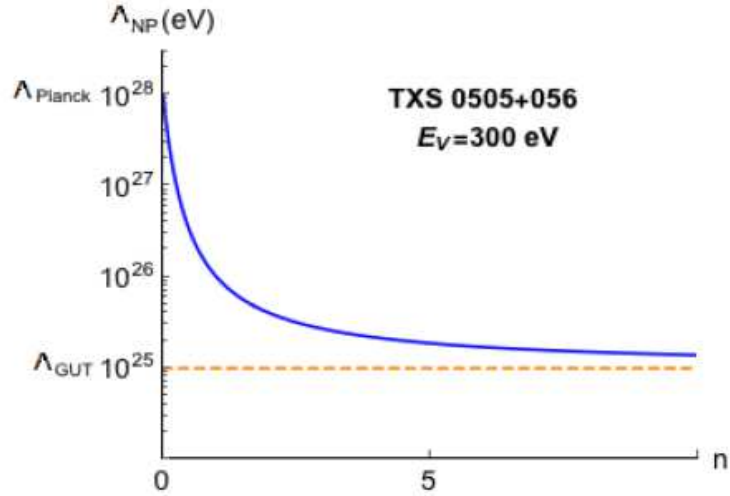}} \vspace*{-.2cm}
\end{center}
\caption{Plot of $\Lambda _{NP}$ vs $n$.} \label{fig1:fig2x}
\end{figure}
\end{center}

At this level, we wish to give one comment. It follows form these
data, including the relation (\ref{eq19}), that we can read two
possible higher bounds for the NP scale. For instance, the Planck
limit bound $10^{28}eV\sim \Lambda _{Planck}$ is recover by taking
$n=0$, being a natural high scale limit associated with string
theory \cite{20}. It is recalled that this theory of  quantum
gravity involves many fields obtained from closed and open string
models providing a neutrino rich sector. It could offer candidates
for the present discussion of HEN. It is worth noting that in the
higher bound corresponding to large values of $n$, we get
\begin{equation}
\Lambda _{NP}\rfloor _{n=\infty}\lesssim 10^{25}eV\sim \Lambda
_{GUT} \label{eq20}
\end{equation}%
being of the order of the GUT scale $\sim 10^{16}GeV$. This theory
involves
a large gauge symmetry going beyond the SM one. It has been shown that $%
SO(10)$ is the simplest gauge group used in the discussion of
neutrino masses. One may expect that such bigger gauge symmetries
could play a relevant r\^{o}le in the elaboration of NP. It is
suggested that the underlying NP seems to be apparently beyond the
range of the recent experiments $\sim TeV$. It turns out  that
experiments will continue to probe higher energies and results
become more accurate. in this way, a  possibility of detecting NP
signatures, including  neutrino mass mechanism, new particles, new
interactions and so on, remains an investigation potential in
future.

\section{Discussion and perspectives}

We have shown that $TeV$-$PeV$ HENs  from cosmic sources can be used
as a useful tool for particle physics effective field theories. This
could  provide  a framework where one can derive stringent bounds
and effects of the NP on the low-energy theory. After the detection
of the HEN from the  blazar TXS 0506+056 by the IceCube-170922A,
along with the successful performance of Fermi-LAT, MAGIC, and the
quick follow-up observations by the other collaborators, significant
progresses have been made recently which could be explored and
interpreted consistently in future.

In this work, we have exploited the IceCube-170922A data to
investigate the NP effect related to neutrinos at low-energies.  Precisely, we have considered the HEN of the TXS 0506+056 with the
SM neutrino data to suspect the scale of the underlying $\Lambda
_{NP}$. Despite our analysis is based on approximate methods, but
along with the known data, we have been able to probe the underlying
NP. Using the results derived from the performed analysis in terms
of the neutrino kinematic with respect to the photon, we have
bounded the underlying NP related the neutrino mass $\Lambda
_{NP}\lesssim \Lambda _{GUT}$.

While neutrinos have underlined much of our thinking about NP from
high energy particle cosmic sources, we still believe that neutrinos
have not yet unveiled their surprising faculties revising our
understanding of nature. More neutrino astonishments are expected in
the few coming years.

\textbf{Acknowledgements}: The authors are grateful to their
families for support.

\end{document}